\documentclass[12pt]{article}
\usepackage[margin=1in]{geometry} 
\usepackage{mathrsfs}
\usepackage{amssymb,amsmath}
\usepackage{amsthm}
\usepackage{bbold} 
\usepackage{physics} 
\usepackage{gensymb} 
\usepackage{float} 
\usepackage[caption=false]{subfig}
\usepackage{hyperref} 
\hypersetup{colorlinks=true,linkcolor=blue,urlcolor=blue,citecolor=blue}
\usepackage{enumerate}
\usepackage{graphicx}
\usepackage{adjustbox}
\usepackage{diagbox} 
\usepackage{color}
\usepackage{caption} 
\usepackage{subcaption}
\usepackage[UKenglish]{babel}

\usepackage{authblk}

\usepackage{tikz}
\usetikzlibrary{shapes.geometric, arrows,arrows.meta}
\begin{document}
\title{Emergent boundary-memory from unitarity constraints in a minimal two interacting quantum particles}
\date{}
	
\author[1]{Ibrahim Yahaya Muhammad}
\author[1,2]{Tanapat Deesuwan}
\author[1]{Monsit Tanasittikosol}
	
\affil[1]{Theoretical and Computational Physics (TCP) Group, Department of Physics, Faculty of Science, King Mongkut's University of Technology Thonburi, Bangkok, 10140, Thailand}
\affil[2]{Quantum Computing and Information Research Centre (QX), Faculty of Science, King Mongkut’s University of Technology Thonburi, Bangkok, 10140, Thailand}
	
\renewcommand\Authands{, } 
\renewcommand\Affilfont{\small} 
	
\maketitle

\abstract
We construct a simple model of two particles mutually attracting/repulsing in a one-dimensional lattice and investigate the corresponding dynamics. We show that enforcing reversible unitary evolution on a minimal direct-motion interaction rule necessarily gives rise to emergent boundary-memory and non-trivial dynamics. Remarkably, the resulting boundary-memory leads to event-horizon-like behaviour, entanglement between subsystem, and Page-curve-like entanglement dynamics.
\section{Introduction}
Dynamics of interacting quantum systems play a central role in physics and underpin a wide range of emerging quantum technologies \cite{cirac2012goals,preskill2018quantum}. Despite their importance, interacting quantum systems are notoriously difficult to analyse due to the rapid growth of complexity with system size, which often scales exponentially with the number of constituents \cite{schuch2009computational}. Consequently, considerable effort has been devoted to the study of minimal settings, such as systems of two interacting particles, which already capture essential features of interaction-driven dynamics. Investigations of two-particle systems have provided fundamental insight into many-body phenomena and have found applications in quantum simulation and condensed matter physics \cite{bloch2012quantum,blume2012few,bloch2008many,georgescu2014quantum}.

To explore such systems, simplified and controllable computational frameworks are often employed. One such framework is quantum cellular automata (QCA), which provides a discrete, local, and unitary description of quantum dynamics on a lattice \cite{arrighi2019overview,farrelly2020review}. Within this broad class, quantum walks (QWs) can be viewed as a particular realisation of QCA, distinguished by an explicit decomposition of the dynamics into internal and translational degrees of freedom \cite{Aharonov1993}. Originally introduced as quantum analogues of classical random walks, QWs exhibit characteristic quantum features such as superposition and interference, leading to faster spreading compared to their classical counterparts. These properties have made QWs a powerful tool for quantum search algorithms \cite{Shenvi2003,portugal2013,childs2014spatial}, as well as for applications in quantum computation \cite{Childs2013b,Singh2021,gong2021}, quantum communication \cite{wang2017,chen2019,Giordani2021}, and high-energy physics \cite{bepari2022,angles2022,sahu2024open}. A wide range of physical implementations further highlights their relevance for future quantum technologies \cite{wang2013,Preiss2015,mugel2016,flurin2017,Zhou2019}, and several comprehensive reviews of QWs are available in the literature \cite{Kempe2003,venegas2012,Xia2020,Kadian2021,qiang2024}.

QWs are generally classified into continuous-time and discrete-time formulations. In this paper, we focus on discrete-time quantum walks (DTQW). The two-particle DTQW was first introduced by Omar \emph{et al.} \cite{Omar2006}, who demonstrated that symmetry properties of the initial state can strongly influence the evolution of two particles, even in the absence of explicit interactions. Subsequent studies incorporated interactions, enabling the simulation of a wide range of quantum phenomena. For example, Wang \emph{et al.} showed that interactions can significantly affect particle behaviour depending on quantum statistics, with bosons exhibiting effective repulsion and fermions forming bound states \cite{Wang2016}. Interactions between two bosonic or fermionic particles have also been shown to generate stable bound states relevant to molecule formation \cite{Ahlbrecht2012,Alonso-Lobo2018}. In addition, interacting two-particle QWs have been explored as a resource for entanglement generation and enhancement, as well as for addressing computational problems such as graph isomorphism \cite{Badhani2021,Carson2015,Berry2011}.

Entanglement dynamics and information scrambling have been central topics in modern studies of quantum systems. Models such as random unitary circuits and toy many-body systems have been used to explore aspects of information scrambling and Page-curve-like dynamics \cite{lau2022page,kehrein2024page,glatthard2024page,saha2024generalized,jha2025page,glatthard2025thermodynamics,ganguly2025quantum,ray2025page}. However, to the best of our knowledge, DTQW models have not previously been employed to investigate entanglement dynamics for two quantum particles interacting through motion-based rules. We bridge this gap by constructing and analysing a minimal model of two interacting particles undergoing a DTQW.

In this work, we construct and study a minimal model of two interacting particles undergoing a DTQW on a one-dimensional lattice, formulated within the QWs framework and viewed from a broader QCA perspective. The interaction is spin-dependent: particles with identical spins move towards each other, while particles with opposite spins move away from each other, with both particles shifting by one lattice site at each time step. While interacting DTQW are commonly implemented via coin- or position-dependent phase factors that encode interactions as local phase accumulation on an otherwise fixed geometry \cite{li2013analysis,li2013discrete,sun2018interacting,verga2018entanglement}, our model instead implements interactions exclusively through conditional shift operations. The coin operation is kept trivial, and no interaction-induced phases are introduced. This modelling choice allows us to isolate interaction effects arising from constrained kinematics and information flow, rather than from phase-based interference. We show that implementing direct motion-based interactions in one dimension generically produces many-to-one evolution near short separations, thereby obstructing reversibility. Imposing unitarity under these constraints necessitates the introduction of additional structural features which we call boundary-memory. The emergence of boundary-memory leads to dynamical behaviour qualitatively resembling event-horizon formation, entanglement between subsystem and Page-curve-like entanglement dynamics, though our model is not black-hole model and has no gravitational effects or relativistic features.

The paper is organised as follows. In Section \ref{sec2}, we introduce our simple model of two interacting particles and construct the unitary evolution operator. In Section \ref{sec3}, we discuss our findings. Finally, we provide the conclusion of our paper in Section \ref{sec4}.
%
\section{\label{sec2}Model and unitary operator}
Our model consists of two interacting particles on a one-dimensional lattice. In contrast to other models that implicitly incorporate interactions by adding local phases to induce interference \cite{Wang2016,Carson2015}, our approach attempts to realise interaction directly through kinematic motion rules.
\subsection{Interaction rule}
Let us define the Hilbert space of the system as $\mathcal{H} = \mathcal{H}_c \otimes \mathcal{H}_p$, where the spin Hilbert space $\mathcal{H}_c$ consists of four orthonormal basis states, $\{\ket{\uparrow\uparrow}, \ket{\uparrow\downarrow}, \ket{\downarrow\uparrow}, \ket{\downarrow\downarrow}\}$ and the Hilbert space of the position state $\mathcal{H}_p$ is spanned by the basis state $\ket{x_1,x_2}$, where $x_1,x_2$ $\in$ $\mathbb{Z}$.

We begin by attempting to construct a \emph{simple} and \emph{consistent} model of interacting quantum systems with the following features and rules:
\begin{enumerate}[i.]
    \item $(1+1)$-dimensional lattice model
    \item spin-parity-dependent global unitary dynamics (shift operator) :  \label{unitary requirement}
    \begin{enumerate}
        \item different spins $\rightarrow$ move further (``repulsion'')
        \item same spins, different positions $\rightarrow$ move closer (``attraction'')
        \item same spins, same positions $\rightarrow$ no motion.
    \end{enumerate}
\end{enumerate}
The spin-parity-dependent part serves as a simple interacting mechanism. We can see immediately that the interaction respects \emph{exchange symmetry} because it depends only on the relative spin parity rather than on the individual spins. By restricting the system to a single spatial dimension on a lattice, we avoid complicated geometry. The need for the global unitary dynamics is to ensure that the model represents a total closed system where information is strictly conserved and there is no decoherence. Given these features and rules, the central question becomes whether a consistent
unitary evolution can be constructed and, if so, what form the corresponding operator must take.
\subsection{\label{sec2.2}Breakdown of unitarity}
As an initial attempt, we propose a simple interaction operator. For simplicity, we choose not to include the term that is responsible for the independent motion of each particle, keeping only the interaction part that conforms with Requirement \ref{unitary requirement}. Also, for the sake of simplicity, we assume that the speed of each particle is always constant at one lattice site per time step, if it moves.

The interaction operator is as follows:
\begin{eqnarray}
	\hat{\mathcal{O}}_{int, naive} &=& (\ket{\uparrow \uparrow}\bra{\uparrow \uparrow} + \ket{\downarrow \downarrow} \bra{\downarrow \downarrow})\otimes \Bigl( \sum_{x_1 > x_2} \ket{x_1-1, x_2+1}\bra{x_1, x_2} \nonumber \\
	&&+ \sum_{x_1 < x_2} \ket{x_1+1, x_2-1}\bra{x_1, x_2}  
    + \sum_{x_1 = x_2} \ket{x_1, x_2}\bra{x_1, x_2} \Bigr) \nonumber \\
    &&+ (\ket{\uparrow \downarrow} \bra{\uparrow \downarrow} + \ket{\downarrow \uparrow} \bra{\downarrow \uparrow}) \otimes \Bigr[ \sum_{x_1 > x_2} \ket{x_1+1, x_2-1}\bra{x_1, x_2} \nonumber \\
	&&+ \sum_{x_1 < x_2} \ket{x_1-1, x_2+1}\bra{x_1, x_2} + \frac{1}{\sqrt{2}}\sum_{x_1=x_2}(\ket{x_1-1, x_2+1}\bra{x_1, x_2} \nonumber \\
    &&+\ket{x_1+1, x_2-1}\bra{x_1, x_2}) \Bigr]
\end{eqnarray}
This operator moves identical-spin particles closer together and pushes opposite-spin particles apart if the two particles are not at the same point initially. If the two particles are at the same point, then they will remain there. Note that the operator does not affect the phase. The interaction is only implemented by a set of rules that govern how particles move between lattice points directly, rather than changing their phases, and let the interference change the probability of finding the particles. Another point to note is that we do not include the coin operators here. This is due to the exchange symmetry implied by the chosen interaction, which dictates that the coin operators can actually be included, but they must act identically on both sides. But, since the interaction depends only on the parity of the spins, the simultaneous changes of both spins cannot change the interaction. Therefore, we can opt not to include them in the first place to make the form as simple as it can be. With these properties, the operator above appears to satisfy all the required conditions. However, this operator is not yet unitary because the map becomes non-invertible in the attractive case when the two particles are too close, within a range where their positions may overlap or swap.

The naive operator can be shown to be non-unitary in several ways. Here, we present the simplest demonstration using a concrete example. Consider the initial state
\begin{eqnarray}
    \ket{\psi}_{0}=\ket{\uparrow\uparrow}_{c}\otimes\ket{1,3}_p
\end{eqnarray}
where the two particles have identical spins and are initially located at lattice positions 1 and 3, respectively. Applying the naive evolution operator once gives
\begin{eqnarray}
    \ket{\psi}_{1}=\ket{\uparrow\uparrow}_{c}\otimes\ket{2,2}_p
\end{eqnarray}
However, the same final state is obtained if the system instead starts from the initial state
\begin{eqnarray}
    \ket{\psi}_{0}=\ket{\uparrow\uparrow}_{c}\otimes\ket{2,2}_p
\end{eqnarray}
since this state remains unchanged under the naive evolution rule. Thus, two distinct initial states evolve into the same final state, implying that the evolution is many-to-one. Consequently, the inverse evolution is not uniquely defined, demonstrating that the naive operator cannot be unitary.

More generally, the naive operator fails to be unitary whenever the separation between the two particles is less than or equal to two lattice sites, since the attractive dynamics becomes many-to-one in this regime. The dynamics therefore naturally partitions the lattice into two regions: the \textbf{``internal region''}, where the particle separation is less than or equal to two lattice sites, and the \textbf{``external region''}, where the separation exceeds two lattice sites. The emergence of the internal region necessitates a modification of the evolution operator. The simplest approach is to alter only the interaction rule at the boundary of the internal region, preventing the particles from entering it while leaving the dynamics in the external region unchanged. Without breaking exchange symmetry, there are only two possible choices: either the particles are reflected back into the external region or they remain stationary at the boundary. However, neither modification restores unitarity. A straightforward examination shows that both still lead to non-invertible evolution. This motivates the introduction of an additional degree of freedom, which provides a minimal extension of the Hilbert space that restores reversible unitary evolution.

\subsection{Emergent boundary-memory} 
All the approaches tried in \ref{sec2.2} fail for the same fundamental reason: the lack of sufficient degrees of freedom necessary to achieve a unitary evolution that satisfies all required conditions. To illustrate this clearly, we can compare a unitary transformation to a rotation, where each input state uniquely corresponds to exactly one output state. However, when two particles converge towards each other along a one-dimensional path due to identical spins (attraction), it is evident that several positions near their midpoint will inevitably map multiple input states to a single output state. In other words, multiple initial points converge to a common endpoint, violating the unitarity requirement of a one-to-one mapping. This issue persists even if we modify the interaction within the internal region, making it either repulsive or stationary. Moreover, a similar issue arises for repulsive interactions if the lattice is finite. To overcome this limitation, a straightforward approach is to introduce at least one additional degree of freedom that acts as a memory, effectively enabling the trajectory to "shift" into another dimension and restoring the one-to-one correspondence necessary for unitarity. The proposed additional degree of freedom is not a unique solution, but rather one of the simplest unitary completions. The specific evolution of this auxiliary degree of freedom is not unique; here, we adopt the simplest implementation. The memory may be finite or infinite with the following properties:
\begin{enumerate}
\item \textbf{Infinite memory}: When two attracting particles enter the internal region, they remain frozen indefinitely while the memory continuously increments like a clock. 
\item \textbf{Finite memory}: When two attracting particles enter the internal region, their motion halts, and the memory begins counting. Once the memory reaches its maximum state, the particles are ejected from the region, after which the interaction effectively switches from attraction to repulsion without altering the spins. The memory then remains fixed in its final state.
\end{enumerate}

\begin{figure}[H]
	\centering
	\subfloat[]{
		\begin{tikzpicture}[>=Stealth, thick, scale=0.6]
			\draw[->] (0,0) -- (0,9.3);
			\node[rotate=90] at (-0.5,4) {\textbf{Time step}};
			\draw[->] (0,0) -- (8.3,0);
			\node[anchor=north west] at (2.5,-1) {\textbf{Position}};
            \foreach \x in {0,1,2,3,4,5,6,7,8} {
            \fill[red] (\x,0) circle (3pt);
            }		
			\draw[dotted, thick] (3,0) -- (3,9);
			\draw[dotted, thick] (5,0) -- (5,9);			
			\node at (1.5,-0.5) {\tiny External};
            \node at (1.5,-0.8) {\tiny region};
			\node at (4,-0.5) {\tiny Internal};
            \node at (4,-0.8) {\tiny region};
			\node at (6.5,-0.5) {\tiny External};
            \node at (6.5,-0.8) {\tiny region};			
			\foreach \x/\y in {0.1/0.1, 1/1, 2/2, 3/3, 5/5, 6/6, 7/7, 8/8} {
				\draw[->, blue] (\x,\y) -- (\x,\y+0.8);
			}
			\foreach \x/\y in {0.1/8, 1/7,  2/6, 3/5, 5/3, 6/2, 7/1, 8/0.1} {
				\draw[->, blue] (\x,\y) -- (\x,\y+0.8);
			}
			\draw[dashed, thick] (3,3.8) -- (5,3.8);
			\draw[dashed, thick] (3,4.9) -- (5,4.9);
			\node at (3.1,4.8) [below right] {\tiny boundary};
            \node at (3.1,4.5) [below right] {\tiny memory};
		\end{tikzpicture}
	}
	\hspace{0.6cm} 
	\subfloat[]{
		\begin{tikzpicture}[>=Stealth, thick, scale=0.6]
			\draw[->] (0,0) -- (0,9.3);
			\node[rotate=90] at (-0.5,4) {\textbf{Time step}};
			\draw[->] (0,0) -- (8.3,0);
			\node[anchor=north west] at (2.5,-1) {\textbf{Position}};
            \foreach \x in {0,1,2,3,4,5,6,7,8} {
            \fill[red] (\x,0) circle (3pt);
            }
			\draw[dotted, thick] (3,0) -- (3,9);
			\draw[dotted, thick] (5,0) -- (5,9);
			\node at (1.5,-0.5) {\tiny External};
            \node at (1.5,-0.8) {\tiny region};
			\node at (4,-0.5) {\tiny Internal};
            \node at (4,-0.8) {\tiny region};
			\node at (6.5,-0.5) {\tiny External};
            \node at (6.5,-0.8) {\tiny region};
			\foreach \x/\y in {4.1/0.1, 5/1, 6/2, 7/3, 8/4} {
				\draw[->, blue] (\x,\y+0.8) -- (\x,\y);
			}
			\foreach \x/\y in {0.1/4, 1/3,  2/2, 3/1, 4/0.1} {
				\draw[->, blue] (\x,\y) -- (\x,\y+0.8);
			}
		\end{tikzpicture}
	}
	\caption{Schematic diagram (a) attracting particles and the boundary-memory, (b) repelling particles. The red dots indicate the lattice points}
	\label{figi0}
\end{figure}
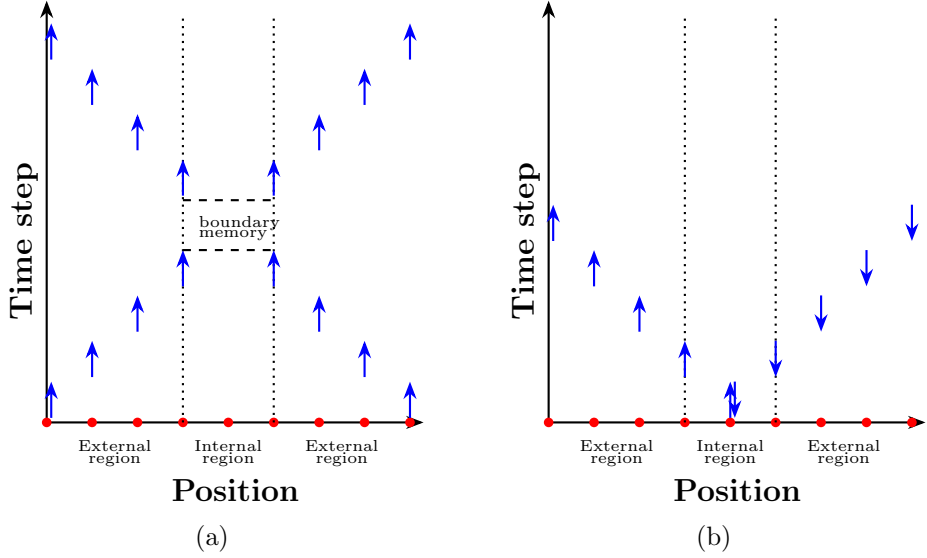

With the additional degree of freedom, we then modified our dynamics for the attracting case to fulfill the unitary requirement as follows. We choose to work with a finite memory for the reason of mapping our model to other physical phenomena later. We refer to this finite memory as \textbf{boundary-memory}. When two particles under attraction enter the internal region, their motion halts immediately. The boundary-memory serves as a counter that incrementally counts up starting from the state $\ket{0}_{b}$. Once this boundary-memory attains its maximum value, the two particles are expelled from the internal region in opposite directions. At this point, their interaction transforms from attraction to repulsion without affecting the spins, as shown in Fig.~\ref{figi0}. We emphasise that the operator remains unitary, and no spin alterations occur during the dynamics.

To show this mathematically, let us start by considering the total Hilbert space of the system with the addition of the boundary-memory, $\mathcal{H} = \mathcal{H}_c \otimes \mathcal{H}_p \otimes \mathcal{H}_{b}$. Where the boundary-memory Hilbert space $\mathcal{H}_{b}$ is spanned by orthonormal basis states $\ket{d}$ where $d \in \{0,1,\dots,n-1\}$ and $n \in \mathbb{Z^+}$. The one-time-step interaction unitary operator is defined as follows:
\begin{eqnarray}
 \hat{U}_{int} &=& P_{same} \otimes \Bigl[ \Bigl( \sum_{x_1 > x_2+2} T_{x_1-1, x_2+1} + \sum_{x_1 < x_2-2} T_{x_1+1, x_2-1}\Bigr) \otimes \ket{0}\bra{0} \nonumber \\
 && + \Bigl( \sum_{x_1 > x_2+2} P_{x_1,x_2} + \sum_{x_1 < x_2-2} P_{x_1,x_2} \Bigr) \otimes \sum_{d=1}^{n-2}\ket{d}\bra{d} \nonumber \\
 &&+ \sum_{x_2-2\leq x_1\leq x_2+2}P_{x_1,x_2} \otimes \sum_{d=0}^{n-2}\ket{d+1}\bra{d} + \Bigl( \sum_{x_1 > x_2} T_{x_1+1, x_2-1} \nonumber \\
 &&+ \sum_{x_1 < x_2} T_{x_1-1, x_2+1} + \frac{1}{\sqrt{2}}\sum_{x_1=x_2}\{T_{x_1-1, x_2+1} + T_{x_1+1, x_2-1}\} \Bigr) \nonumber \\
    &&\otimes \ket{n-1}\bra{n-1} \Bigr] + P_{diff} \otimes \Bigr[ \sum_{x_1 > x_2} T_{x_1+1, x_2-1} + \sum_{x_1 < x_2} T_{x_1-1, x_2+1} \nonumber \\
 &&+ \frac{1}{\sqrt{2}}\sum_{x_1=x_2}(T_{x_1-1, x_2+1} + T_{x_1+1, x_2-1}) \Bigr] \otimes \sum_{d=0}^{n-1} \ket{d}\bra{d}
 \label{ei2}
\end{eqnarray}
where $P_{same}=\ket{\uparrow \uparrow}\bra{\uparrow \uparrow} + \ket{\downarrow \downarrow} \bra{\downarrow \downarrow}$, $P_{diff}=\ket{\uparrow \downarrow}\bra{\uparrow \downarrow} + \ket{\downarrow \uparrow} \bra{\downarrow \uparrow}$, $P_{x_1,x_2}=\ket{x_1,x_2}\bra{x_1, x_2}$, $T_{x_1-1,x_2+1}=\ket{x_1-1, x_2+1}\bra{x_1, x_2}$, and $T_{x_1+1,x_2-1}=\ket{x_1+1, x_2-1}\bra{x_1, x_2}$.
%
\section{\label{sec3}Entanglement}
In constructing the unitary operator governing the dynamics of our model, as described in Section \ref{sec2}, we observed that two distant particles with identical spins in the up-down basis begin to attract each other as expected. Upon approaching the internal region and entering it, the particles become immobilised, effectively trapped. The boundary where particles abruptly become stationary qualitatively resembles an \textbf{event-horizon-like} structure in our model. Also, the emergence of the boundary-memory in our model motivates a deeper investigation into how it can correlate with the other subsystem and also the entanglement dynamics.

\subsection{\label{sec3.1}Entanglement between subsystems}
\subsubsection{Superposition and localised positions}
Consider a case where one particle is in superposition on the left and the other is localised at a point on the right.
\\
\\
\textit{Case I: Odd separation distance}: If the two particles have odd separation distances with initial state given by
\begin{eqnarray}
 \ket{\psi}_{0}=\frac{1}{\sqrt{2}}\ket{\uparrow\uparrow}_{c}\otimes(\ket{-2,2}+\ket{-1,2})_p \otimes \ket{0}_{b}.
\end{eqnarray}
The state after applying the unitary once is
\begin{eqnarray}
 \ket{\psi}_{1}&=&\frac{1}{\sqrt{2}}\ket{\uparrow\uparrow}_{c}\otimes(\ket{-1,1}+\ket{0,1})_p \otimes \ket{0}_{b}.
\end{eqnarray}
The state after n$^{th}$ step is
\begin{eqnarray}
 \ket{\psi}_{n}&=&\frac{1}{\sqrt{2}}\ket{\uparrow\uparrow}_{c}\otimes(\ket{-1,1}+\ket{0,1})_p \otimes \ket{n-1}_{b}.
\end{eqnarray}
Note that in this state, the boundary-memory does not exhibit a quantum correlation between any of the particles.
\\
\\
\textit{Case II: Even separation distance}: When the two particles have even separation distances, the state is  given by
\begin{eqnarray}
 \ket{\psi}_{0}=\frac{1}{\sqrt{2}}\ket{\uparrow\uparrow}_{c}\otimes(\ket{-3,2}+\ket{-2,2})_p \otimes \ket{0}_{b}.
\end{eqnarray}
The state after applying the unitary once
\begin{eqnarray}
 \ket{\psi}_{1}&=&\frac{1}{\sqrt{2}}\ket{\uparrow\uparrow}_{c}\otimes(\ket{-2,1}+\ket{-1,1})_p \otimes \ket{0}_{b}.
\end{eqnarray}
However, after applying the unitary operator twice, the state becomes
\begin{eqnarray}
 \ket{\psi}_{2}&=&\frac{1}{\sqrt{2}}\ket{\uparrow\uparrow}_{c}\otimes(\ket{-1,0}_p\otimes \ket{0}_{b} + \ket{-1,1}_p \otimes \ket{1}_{b})
\end{eqnarray}
which is an entangled state. Therefore, the two particles become entangled with the boundary-memory.
\subsubsection{Superposition in positions}
Consider the case where one particle is in superposition on the left and the other is also in superposition on the right.
\\
\\
\textit{Case I: Odd separation distance}: The initial state in which the two particles are in superposition in positions with odd separation distance is given by
\begin{eqnarray}
\ket{\psi}_{0}&=&\frac{1}{2}\ket{\uparrow\uparrow}_{c}\otimes(\ket{-2}+\ket{-1})_{p_1}\otimes(\ket{2}+\ket{3})_{p_2} \otimes \ket{0}_{b}. \nonumber \\
 &=&\frac{1}{2}\ket{\uparrow\uparrow}_{c}\otimes(\ket{-2,2}+\ket{-2,3}+\ket{-1,2}+\ket{-1,3})_p \otimes \ket{0}_{b}.
\end{eqnarray}
The state after applying the unitary once is
\begin{eqnarray}
 \ket{\psi}_{1}&=&\frac{1}{2}\ket{\uparrow\uparrow}_{c}\otimes(\ket{-1,1}+\ket{-1,2}+\ket{0,1}+\ket{0,2})_p \otimes \ket{0}_{b}.
\end{eqnarray}
Applying the unitary twice gives the following result.
\begin{eqnarray}
 \ket{\psi}_{2}&=&\frac{1}{2}\ket{\uparrow\uparrow}_{c}\otimes\big[ (\ket{-1,1}+\ket{0,1}+\ket{0,2})_p \otimes \ket{1}_{b} +\ket{0,1}_{p} \otimes \ket{0}_{b}\big].
\end{eqnarray}
We can see that the particles and the boundary-memory become entangled. If we post-select on the basis of the boundary-memory $\ket{1}\bra{1}_{b}$, the two particles become entangled. This demonstrates that boundary-memory can serve as a mechanism for generating entanglement between initially uncorrelated particles.
\\
\\
\textit{Case II: Even separation distance}: The state is  given by
\begin{eqnarray}
\ket{\psi}_{0}&=&\frac{1}{2}\ket{\uparrow\uparrow}_{c}\otimes(\ket{-2}+\ket{-1})_{p_1}\otimes(\ket{3}+\ket{4})_{p_2} \otimes \ket{0}_{b}. \nonumber \\
 &=&\frac{1}{2}\ket{\uparrow\uparrow}_{c}\otimes(\ket{-2,3}+\ket{-2,4}+\ket{-1,3}+\ket{-1,4})_p \otimes \ket{0}_{b}.
\end{eqnarray}
The state after applying the unitary once is
\begin{eqnarray}
 \ket{\psi}_{1}&=&\frac{1}{2}\ket{\uparrow\uparrow}_{c}\otimes(\ket{-1,2}+\ket{-1,3}+\ket{0,2}+\ket{0,3})_p \otimes \ket{0}_{b}.
\end{eqnarray}
Applying the unitary twice gives the following result.
\begin{eqnarray}
 \ket{\psi}_{2}&=&\frac{1}{2}\ket{\uparrow\uparrow}_{c}\otimes\big[ (\ket{0,1}+\ket{0,2}+\ket{1,2})_p \otimes \ket{0}_{b} +\ket{0,2}_{p} \otimes \ket{1}_{b}\big].
\end{eqnarray}
In this case, the two particles and the boundary-memory become entangled. Also, when we post-select using $\ket{0}\bra{0}_{b}$ basis, we observe that particle $1$ and particle $2$ become entangled.
\subsection{\label{sec3.2}Entanglement dynamics}
We now investigate how the entanglement entropy between subsystems evolves during the dynamics. The total system consists of two particles (subsystem 1) and the boundary-memory (subsystem 2).

Entanglement entropy for this bipartite system is quantified using the von Neumann entropy, defined as
\begin{eqnarray}
	E(\ket{\psi}) = S(\rho_1) = -\text{Tr}(\rho_1 \log_2\rho_1), 
	\label{eqi9}
\end{eqnarray}
where $\rho_1 = \text{Tr}_2(\rho_{12})$ is the reduced density matrix for subsystem 1.

We will analyse four different cases of initial settings and discuss our findings.
\subsubsection{Localised position}
First, consider a situation where two localised particles are positioned on opposite sides, one on the left and the other on the right. The initial state is
\begin{eqnarray}
	\ket{\psi}_{0}=\ket{\uparrow\uparrow}_{c}\otimes\ket{-2,2}_p \otimes \ket{0}_{b}.
\end{eqnarray}
After applying the unitary once, the particles fall into the internal region and freeze. After applying the unitary operator three times, the state evolves to:
\begin{eqnarray}
	\rho_{3}=\ket{\uparrow\uparrow}\bra{\uparrow\uparrow}_{c}\otimes\ket{-1,1}\bra{-1,1}_p \otimes \ket{2}\bra{2}_{b}.
\end{eqnarray}
Tracing out the particles subsystem yields the reduced density matrix of the boundary-memory
\begin{eqnarray}
	\rho_{3,\text{boundary-memory}}=\mathrm{Tr}_{\text{particles}} \left( \rho_3 \right) = \ket{2}\bra{2}_{b}.
\end{eqnarray}
We observe that the state remains pure and that the entanglement entropy is zero, as shown in Fig.~\ref{figi1}(a). Thus, a fully localised initial state produces no entanglement between the particles and the boundary-memory. In this configuration, once the particles become frozen within the internal region and no additional superposition of entry times exists, the entropy remains zero throughout the evolution. More generally, entanglement between the particles and the boundary-memory arises only when the initial state contains superpositions associated with different entry times into the internal region.
\subsubsection{Superposition in position}
\begin{figure}[H]
 \centering
 \subfloat[ ]{\includegraphics[width=0.4\textwidth]{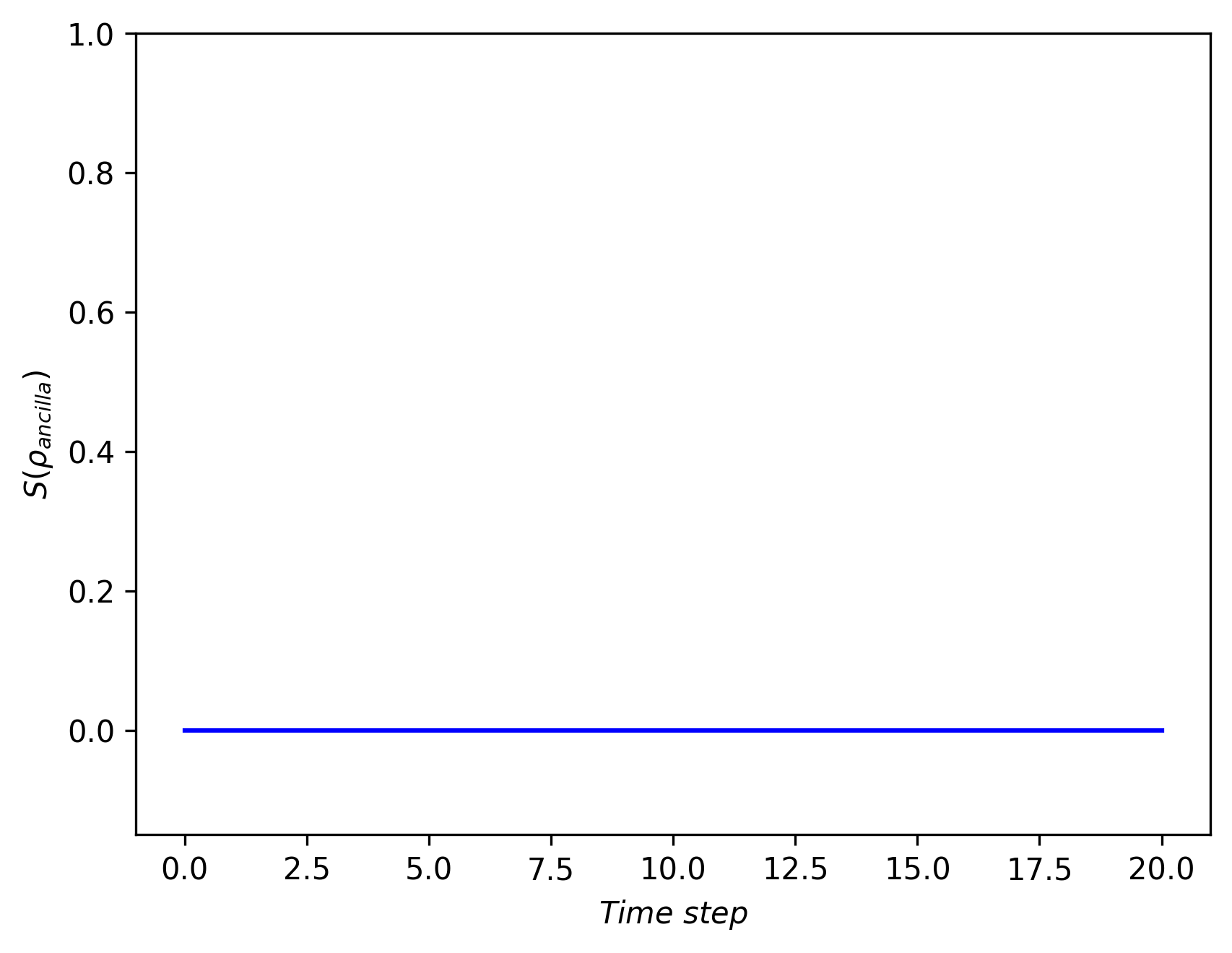}}%
 \hspace{0.2em}
 \subfloat[ ]{\includegraphics[width=0.4\textwidth]{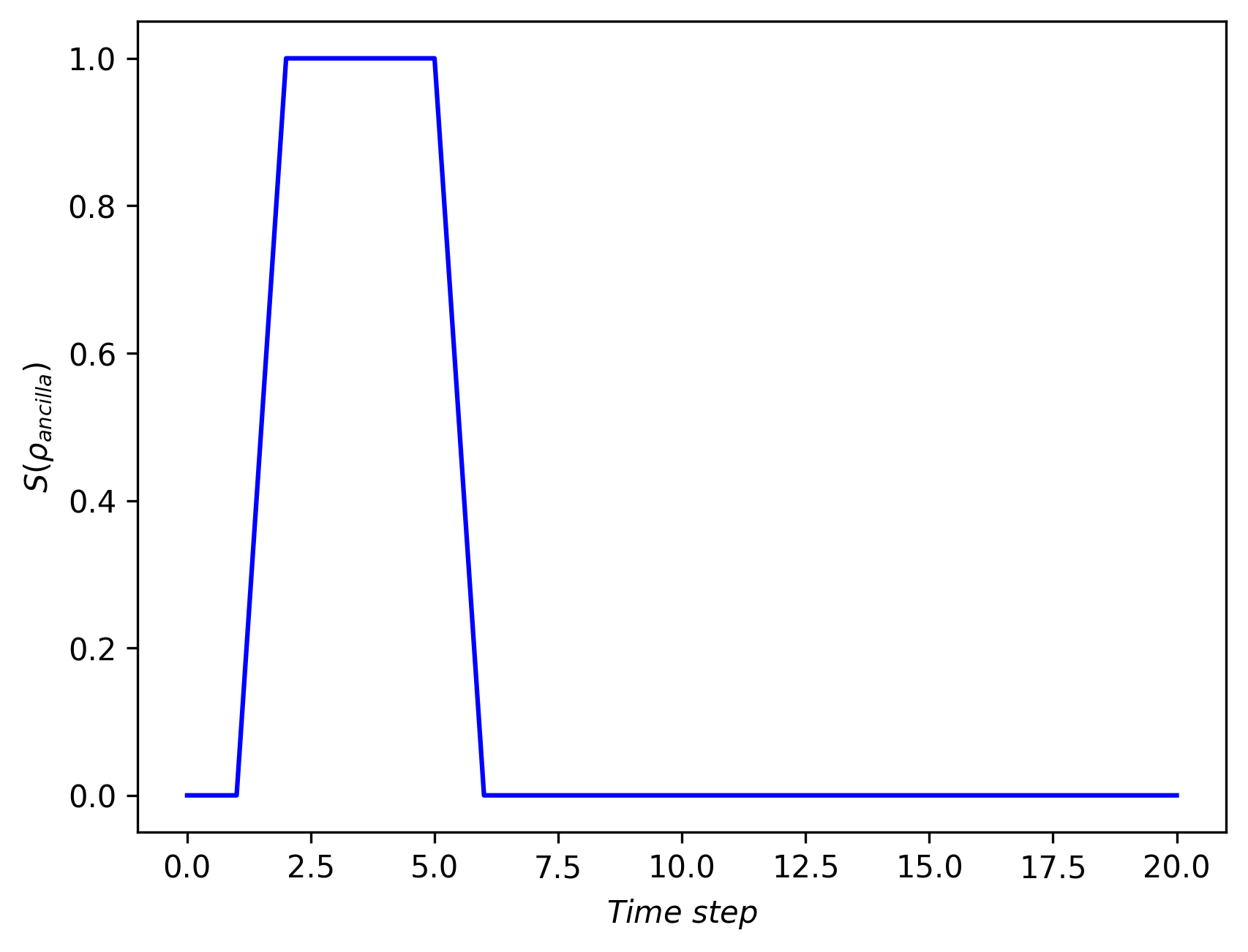}}%
 \caption{Entanglement between the two particles and the boundary-memory subsystem. The initial states considered are (a) localised, (b) superposition in position}
 \label{figi1}
\end{figure}

In contrast to the localised initial state, introducing initial superposition states significantly changes the entanglement dynamics. For instance, an initial state where one particle is in superposition on the left and the other is localised at a point on the right, given by
\begin{eqnarray}
 \ket{\psi}_{0}=\frac{1}{\sqrt{2}}\ket{\uparrow\uparrow}_{c}\otimes(\ket{-3,2}+\ket{-2,2})_p \otimes \ket{0}_{b}.
\end{eqnarray}
The state after applying the unitary once is
\begin{eqnarray}
 \ket{\psi}_{1}&=&\frac{1}{\sqrt{2}}\ket{\uparrow\uparrow}_{c}\otimes(\ket{-2,1}+\ket{-1,1})_p \otimes \ket{0}_{b}.
\end{eqnarray}
In the state above, an event-horizon-like structure formed, but the boundary-memory does not start counting yet. If we trace out the particles, the reduced state of the boundary-memory becomes pure, given as
\begin{eqnarray}
 \rho_{1,\text{boundary-memory}}&=& \ket{0}\bra{0}_{b}.
\end{eqnarray}
The entanglement entropy is zero. However, after applying the unitary operator twice, the state becomes
\begin{eqnarray}
 \ket{\psi}_{2}&=&\frac{1}{\sqrt{2}}\ket{\uparrow\uparrow}_{c}\otimes(\ket{-1,0}_p\otimes \ket{0}_{b} + \ket{-1,1}_p \otimes \ket{1}_{b})
 \label{eq10}
\end{eqnarray}
which is an entangled state. We can see that the reduced boundary-memory state is now mixed. Therefore, it shows how a pure state evolves into a mixed state due to entanglement between the particles and the boundary-memory, indicating a non-trivial interaction between the boundary-memory and the two particles. The particles and the boundary-memory become increasingly entangled as the attractive dynamics drive the particles towards the internal region, leading to a growth in entropy. Eventually, as the particles exit the internal region, the entropy decreases and finally returns to zero, as shown in Fig.~\ref{figi1}(b). The resulting entropy profile qualitatively resembles the Page curve \cite{page1993information,page2013time} associated with unitary black-hole evaporation, characterised by an initial increase followed by a subsequent decrease in entanglement entropy. Although the entropy studied here is not equivalent to radiation entropy in gravitational systems, the observed behaviour demonstrates that Page-curve-like dynamics can emerge naturally in constrained unitary systems possessing temporary information trapping and boundary-memory.
\subsubsection{Superposition in spin}
We consider the initial state, in which only one particle’s spin is in superposition. The state is given by
\begin{eqnarray}
 \ket{\psi}_{0}=\frac{1}{\sqrt{2}}(\ket{\uparrow\uparrow}+\ket{\downarrow\uparrow})_{c}\otimes \ket{-2,2}_p \otimes \ket{0}_{b}.
\end{eqnarray}
This initial state also contains a coherent superposition of attracting and repelling interactions. Applying the evolution operator~$\hat U_{int}$ defined in Eq.~\ref{ei2}, we obtain the state after one step as
\begin{equation}
 \ket{\psi_1}=\frac{1}{\sqrt2}\Bigl[\ket{\uparrow\uparrow}_c\otimes\ket{-1,1}_p\otimes\ket{0}_d+\ket{\downarrow\uparrow}_c\otimes\ket{-3,3}_p\otimes\ket{0}_d\Bigr].
\end{equation}
The attracting branch moves inwards, whereas the repelling branch moves outwards. After the second evolution step, the attracting component reaches the internal region, and the two particles stop. The state becomes
\begin{align}
 \ket{\psi_2}=\frac{1}{\sqrt2}&\Bigl[\ket{\uparrow\uparrow}_c\otimes\ket{-1,1}_p\otimes\ket{1}_d+\ket{\downarrow\uparrow}_c\otimes\ket{-4,4}_p\otimes\ket{0}_d\Bigr].
\end{align}
Tracing out the two particles' degrees of freedom, we obtain
\begin{align}
 \rho_{2,\text{boundary-memory}}=&\frac12(\ket{0}\bra{0}_{b}+\ket{1}\bra{1}_{b}).
\end{align}
The entropy is $S(\rho_{2,\text{boundary-memory}})=1$. We observed that the entanglement entropy after $k$ steps is
\begin{equation}
 S(\rho_{k,\text{boundary-memory}})=\begin{cases}
  0, & k<2,\\
  1, & k\ge 2.
 \end{cases}
\end{equation}
We can see that once the attractive branch falls into the region, the particles and the boundary-memory become entangled. As shown in Fig.~\ref{figi2}(a), the entanglement persists even after the particles associated with the attractive branch have exited the internal region and the boundary-memory has reached its maximum state. The entropy rises to a non-zero value and remains constant thereafter. From the perspective of an external observer, information becomes effectively inaccessible once the particles enter the internal region, where it remains encoded in the boundary-memory. This behaviour mimics certain features associated with irreversible information isolation. These results demonstrate that the model provides a useful framework for exploring non-trivial dynamics in constrained unitary quantum systems.
\subsubsection{Superposition in both spins and positions}
\begin{figure}[H]
 \centering
 \subfloat[ ]{\includegraphics[width=0.4\textwidth]{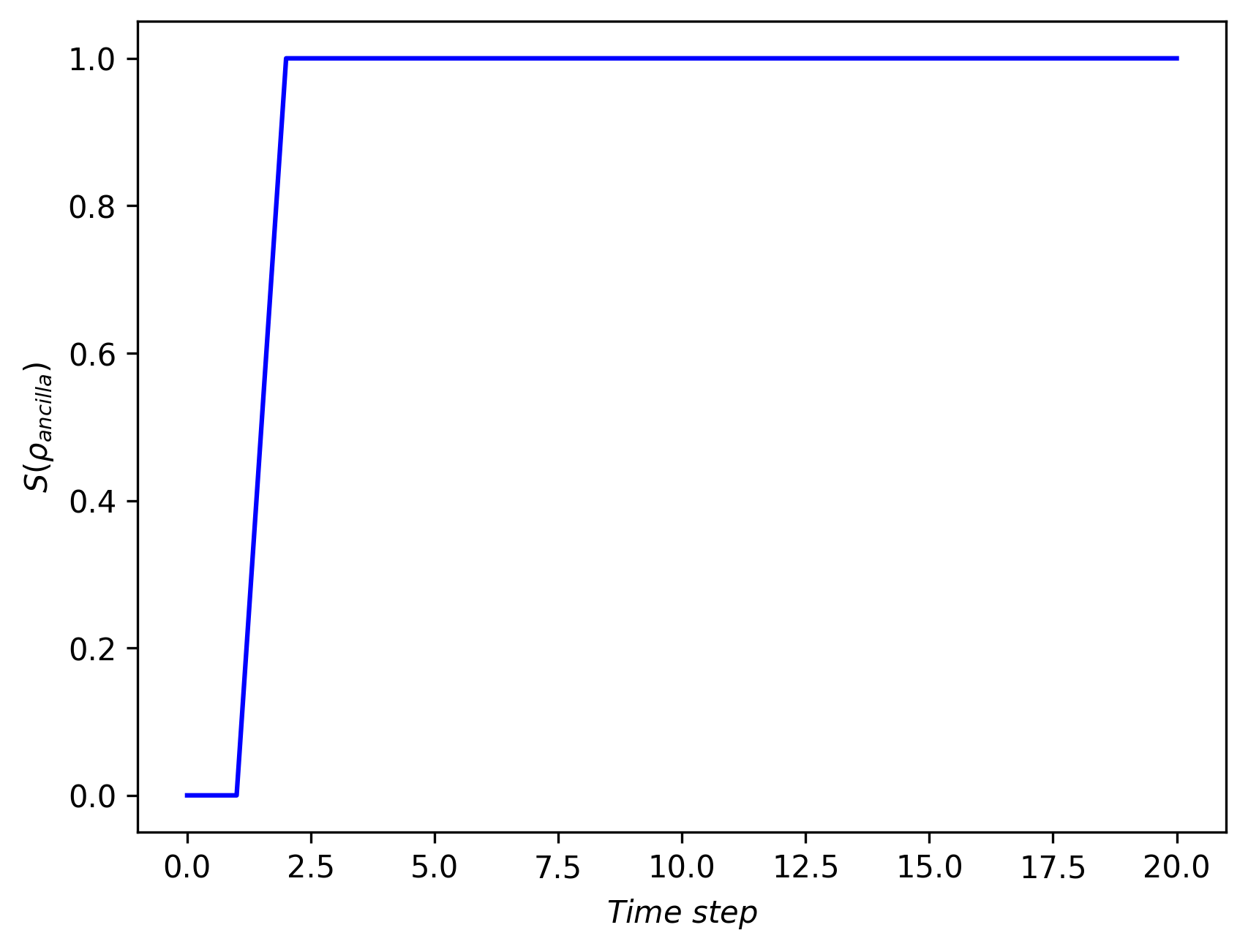}}%
 \hspace{0.2em}
 \subfloat[ ]{\includegraphics[width=0.4\textwidth]{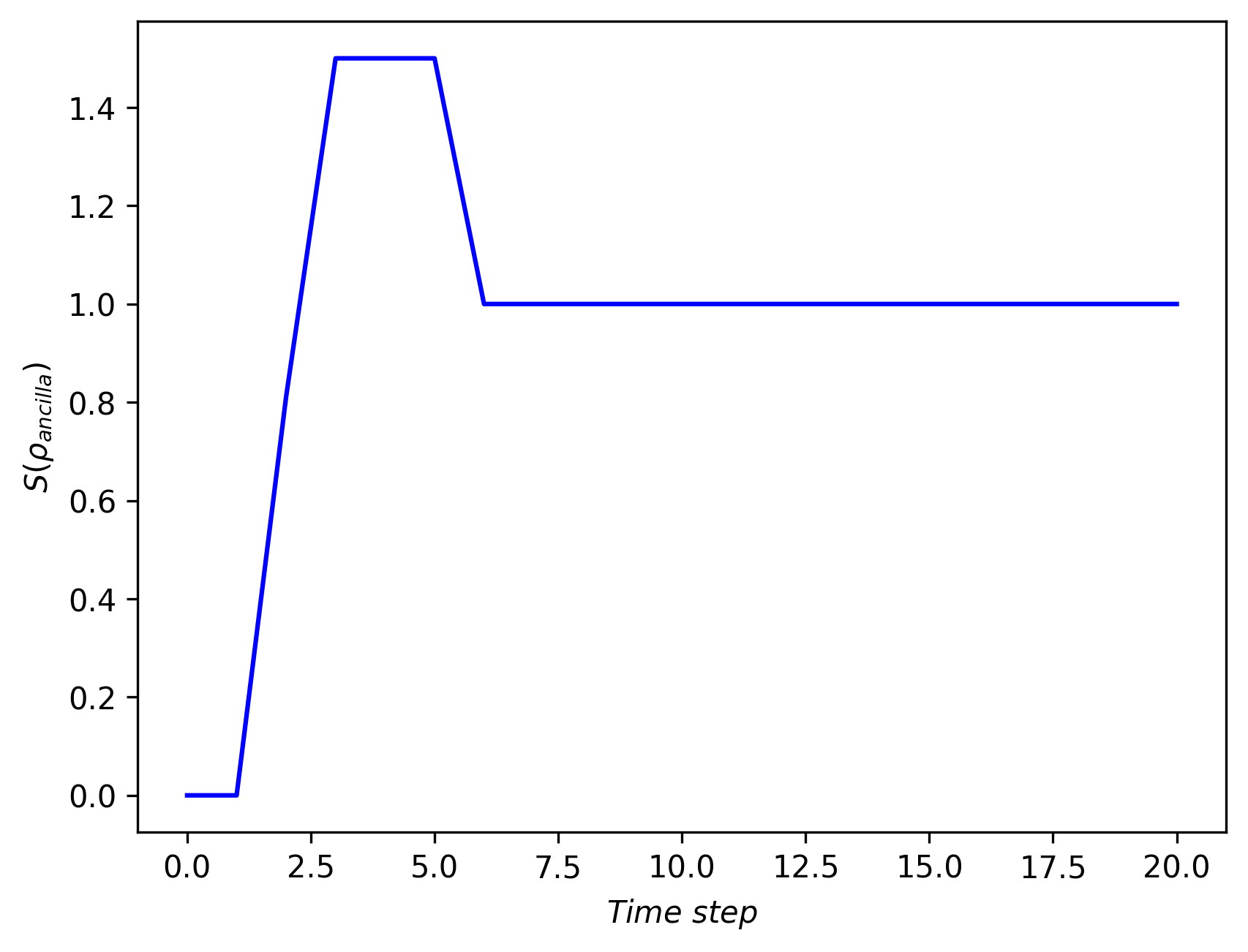}}%
 \caption{Entanglement between the two particles and the boundary-memory subsystem. The initial states considered are in superposition (a) superposition in spins, (b) superposition in both spin and position}
 \label{figi2}
\end{figure}

Further complexity arises when the initial state involves both spins and positions in superpositions. It involves the attraction and repulsion cases; the entanglement in this state may result in interesting dynamics. Let us consider such a setting, where the spin and position of particle one are both in superposition, while the other particle is localised. The initial state is
\begin{eqnarray}
 \ket{\psi}_{0}=\frac{1}{2}(\ket{\uparrow\uparrow}+\ket{\downarrow\uparrow})_{c}\otimes(\ket{-3,2}+\ket{-2,2})_p \otimes \ket{0}_{b}.
\end{eqnarray}
In this state, the two particles simultaneously attract and repel each other. Now, after two evolution steps, the resulting state is
\begin{eqnarray}
 \ket{\psi}_{2}=\frac{1}{2}[\ket{\uparrow\uparrow}_{c}(\ket{-1,0}_p\ket{0}_{b}+\ket{-1,1}_p\ket{1}_{b}) + \ket{\downarrow\uparrow}_{c}(\ket{-5,4}+\ket{-4,4})_p\otimes \ket{0}_{b}]
\end{eqnarray}
and then tracing out the two particles, gives
\begin{eqnarray}
 \rho_{2,\text{boundary-memory}} &=& \frac{3}{4} \ket{0}\bra{0}_{b} + \frac{1}{4} \ket{1}\bra{1}_{b}.
\end{eqnarray}
where the boundary-memory entropy is
\begin{eqnarray}
 S(\rho_{2,\text{boundary-memory}}) &=& 0.811
\end{eqnarray}
The entropy rises to a peak and then decreases as the attracting particles leave the internal region and the boundary-memory reaches its maximum state. In Fig.~\ref{figi2}(b), the observed entanglement-entropy behaviour is not precisely Page-curve-like because the entropy does not return to zero.
\section{\label{sec4}Conclusion}
We constructed and analysed a model of two interacting quantum particles based on a minimal set of assumptions: global unitary dynamics together with spin-parity-dependent attraction and repulsion on a $(1+1)$-dimensional lattice. Unlike conventional interacting quantum walks models, which typically encode interactions through local phase modifications, our model implements interactions directly through conditional particle motion. This distinguishes the nature of the interaction in our framework from previously studied approaches. We found that enforcing strict unitarity on such motion-based interactions leads to an obstruction associated with many-to-one evolution near short separations. Restoring unitarity naturally requires the introduction of an additional degree of freedom, which we termed boundary-memory.

The resulting dynamics exhibit several non-trivial features. In particular, the entanglement entropy between the particles and the boundary-memory displays behaviour qualitatively analogous to Page curves encountered in studies of black-hole evaporation, despite the absence of gravitational or relativistic effects. Furthermore, the localisation of particles at the boundary of the internal region gives rise to an event-horizon-like structure within the dynamics.

Overall, our results demonstrate that even highly simplified interacting two-particle models can exhibit rich dynamical and entanglement behaviour. The model provides a controlled setting for investigating how simple interaction rules, combined with unitary constraints, can influence quantum dynamics. Future work may extend this framework to larger particle numbers, alternative boundary-memory mechanisms, or non-trivial coin operators in order to further explore the relationship between constrained quantum dynamics and emergent information behaviour.
\section{Acknowledgements}
The first author was supported by the ``Petchra Pra Jom Klao Ph.D. Research Scholarship from King Mongkut’s University of Technology Thonburi" Contract No.\,14/2561. This research has received funding support from the NSRF via the Program Management Unit for Human Resources \& Institutional Development, Research and Innovation [grant number B39G680007].

\bibliographystyle{ieeetr}
\bibliography{references_paper}
\end{document}